\documentclass[sigconf,usenames,dvipsnames]{acmart}

\usepackage{dblfloatfix} 
\usepackage{booktabs} 
\usepackage{arydshln}
\usepackage{subfig}
\AtBeginDocument{%
  \providecommand\BibTeX{{%
    \normalfont B\kern-0.5em{\scshape i\kern-0.25em b}\kern-0.8em\TeX}}}

\copyrightyear{2024}
\copyrightyear{2024}
\acmYear{2024}
\setcopyright{acmlicensed}\acmConference[arXiv]{}{May, 2024}{}
\begin{document}
\newcommand{\system}{HoloDevice}
\title{\system{}: Holographic Cross-Device Interactions for Remote Collaboration}

\author{Neil Chulpongsatorn}
\affiliation{%
  \institution{University of Calgary}
  \city{Calgary}
  \country{Canada}}  
\email{thobthai.chulpongsat@ucalgary.ca}

\author{Thien-Kim Nguyen}
\affiliation{%
  \institution{University of Calgary}
  \city{Calgary}
  \country{Canada}}  
\email{thienkim.nguyen@ucalgary.ca}

\author{Nicolai Marquardt}
\affiliation{%
  \institution{Microsoft Research}
  \city{Redmond}
  \country{United States}}  
\email{nicmarquardt@microsoft.com}

\author{Ryo Suzuki}
\affiliation{%
  \institution{University of Calgary}
  \city{Calgary}
  \country{Canada}}
\email{ryo.suzuki@ucalgary.ca}

\renewcommand{\shortauthors}{Chulpongsatorn, Nguyen, Marquardt, and Suzuki}

\begin{abstract}
This paper introduces \textbf{\textit{holographic cross-device interaction}}, a new class of \textit{remote} cross-device interactions between local physical devices and holographically rendered remote devices. Cross-device interactions have enabled a rich set of interactions with device ecologies. Most existing research focuses on \textit{co-located} settings (meaning when users and devices are in the same physical space) to achieve these rich interactions and affordances. In contrast, holographic cross-device interaction allows remote interactions between devices at distant locations by providing a rich visual affordance through real-time holographic rendering of the device's motion, content, and interactions on mixed reality head mounted displays. This maintains the advantages of having a physical device, such as precise input through touch and pen interaction. Through the holographic rendering, not only can remote devices interact as if they are co-located, but they can also be virtually augmented to further enrich interactions, going beyond of what is possible with existing cross-device systems. To demonstrate this concept, we developed \system{}, a prototype system for holographic cross-device interaction using the Microsoft Hololens 2 augmented reality headset. Our contribution is three fold. First we introduces the concept of holographic cross-device interaction. Second we presents a design space containing three unique benefits, which includes: (1) spatial visualization of interaction and motion, (2) rich visual affordances for intermediate transition, and (3) dynamic and fluid configuration. Last we discuss a set of implementation demonstration and use-case scenarios that further explore the space.
\end{abstract}



\begin{teaserfigure}
\includegraphics[width=\textwidth]{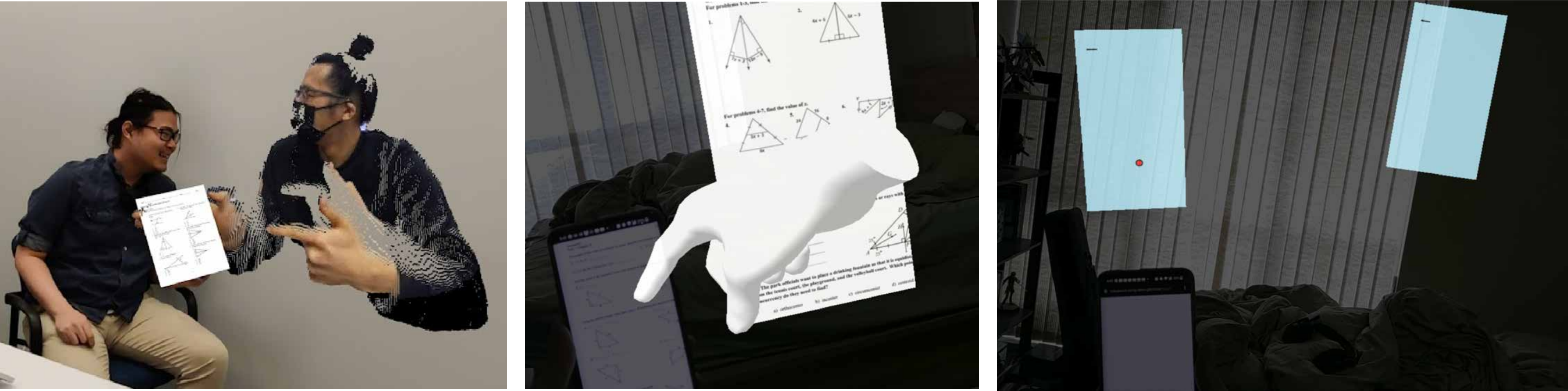}
\caption{We present a mixed reality collaboration space where both users and devices are embodied by a holographic rendering. For example, users can interact with remote collaborators represented by a point cloud along with their device rendered onto the hologram's hand (left), or they can see a 3D model abstraction that interact with the devices in the space (middle). This space also allows remote devices to interact with each other as if they are co-located (right), which opens up a new sector in cross device interactions like never before.}
\label{fig:intro-figure}
\end{teaserfigure}

\maketitle

\section{Introduction}


Cross-device interactions \cite{brudy2019cross}  allow the seamless interactions between multiple, distributed, and inter-connected devices for collaboration \cite{bardram2012reticularspaces, wigdor2009wespace, klokmose2015webstrates}, communication \cite{aumi2013doplink}, and entertainment \cite{baillard2017multi, ajam2017middleware, ballendat2010proxemic}.
As the number of devices is continuously increasing in today's ubiquitous computing era (e.g., phone, tablet, wearable, and desktops, etc) \cite{marquardt2021airconstellations, wang2021gesturar}, designing seamless interactions for this \textit{``device ecology''} become increasingly important, not only in HCI research \cite{brudy2019cross}, but also in commercial products, such as Apple's continuity\footnote{https://www.apple.com/ca/macos/continuity/} and Microsoft's technology in collaboration with Android (the Surface Duo and select Samsung Devices) \footnote{https://support.microsoft.com/en-us/topic/seamlessly-transfer-content-between-your-devices-8a0ead3c-2f15-1338-66ca-70cf4ae81fcb}.
In particular, cross-device interactions enable a set of unique interactions and affordances for \textit{collaboration} by leveraging seamless communication between multiple users and devices.
For example, existing research has demonstrated 
collaborative cross-device interactions for content sharing \cite{xiao2016capcam, klokmose2015webstrates}, photo sorting \cite{lucero2011pass, merrill2007siftables}, and data analysis \cite{badam2016supporting, homaeian2018group}. 

Currently, most research on collaborative cross-device interaction focuses on \textit{co-located} settings---that is, the users and devices need to be in the same physical space to achieve these rich interactions and affordances.
Recent works have began exploring cross-device interactions in remote settings \cite{bardram2012reticularspaces, klokmose2015webstrates}, but they often still lack rich visual and gestural feedback.
For example, a user cannot see how another remote user manipulates the device and content. Rich gestural expressions through embodiment of the user (such as pointing and circling) have shown to be beneficial for coordination in a collaborative setting \cite{tang2010three}. The lack of such embodiment might lead to confusion during collaboration.
Despite the fact that remote collaboration has become more important than ever due to the COVID-19 pandemic, the exploration of collaborative remote cross-device interaction still remains largely unexplored.

In this paper, we introduce \textbf{\textit{holographic cross-device interaction}}, a new approach to enabling \textit{remote} and \textit{collaborative} cross-device interactions.
In contrast to existing remote cross-device interactions, holographic cross-device interaction enables the interaction between local and remote devices through a real-time holographic rendering of the remote device.
This allows the user to see the the remote device's motion, content, and interaction \textit{in real-time}, as if they are co-located in the same physical space.
This real-time holographic rendering enables a variety of existing cross-device interactions in remote collaboration settings. By rendering (parts of) remote participants, expressive gestures can be situated alongside the devices.
Moreover, holographic cross-device interactions also enable a set of unique interaction affordances that are not possible with existing cross-device interactions.
For example, by leveraging the flexibility of virtual rendering, the user can dynamically change the remote device's size, position, content, and alignment in the mixed reality environment. 

To provide a better understanding of this new concept, this paper contributes a taxonomy and interaction design space of holographic cross-device interaction that helps researchers contextualize and explore the domain. We first discuss three unique benefits that designers can take advantage of. This includes (1) spatial visualization of interaction and motion, (2) rich visual affordances for intermediate transition, and (3) dynamic and fluid configuration. Then we present five possible use cases in the space (content transfer, input and control, edit and annotation, adaptive UI, and extended screens). We also list out ten interaction techniques (snapping, automatic alignment, magnetic attachment, spatial relationship, 3D interaction, fluid configuration, explicit connection, visible interaction, history cloning, and scalable collaboration) that can be leveraged for the use cases.     

To demonstrate this concept, we develop \system{}, a prototype system that leverages Microsoft Hololens 2.
We render the device's real-time motion and content with WebXR (A-Frame, AR.js, 8th wall and Three.js) on the Hololens 2, which is captured, shared, and communicated through a WebSocket protocol and sensed through internal accelerometers. 
By leveraging this capability, we demonstrate a variety of interactions  through an implementation of each of the use cases in our design space. We also explore further use of our system by discussing six application scenarios in detail. 
Lastly, we discuss future research opportunities to further improve and address cross-device interaction for remote collaboration. 

In summary, this paper contributes  
\begin{enumerate}
\item the concept of holographic cross-device interaction
\item a taxonomy and interaction design space for holographic cross-device interaction
\item a demonstration of a variety of interactions and applications through \system{} and further exploration through discussions of application scenarios. 
\end{enumerate}
\section{Related Work}

\begin{figure}[h!]
\centering
\includegraphics[width=0.8\linewidth]{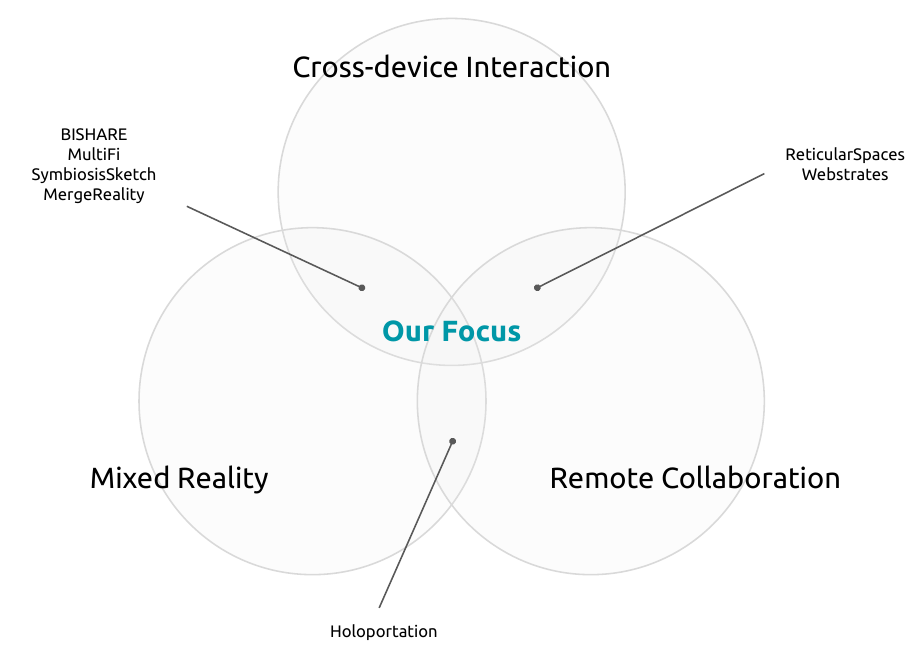}
\caption{Diagram presents where this work is focused on. Which is the intersection of cross-device interaction, remote collaboration, and holographic mixed reality.}
\label{fig:related-work-figure}
\end{figure}

We situate our work in the intersection of \textit{cross-device interaction}, \textit{remote collaboration}, and \textit{holographic mixed reality}.
While there are previous research projects that explore the intersection of two of these aspects, we argue that, to the best of our knowledge, there is no prior work that explore the intersection of all of these three elements (Figure \ref{fig:related-work-figure}). 
In the following sections, we describe the related works in these domains.

\subsection{Cross-Device Interaction with Mixed Reality}
Cross-device interaction is a vast research area that has been investigated over the last decades. 
Cross-device taxonomy survey \cite{brudy2019cross} provides a nice overview of the landscape in the cross-device interaction research, summarizing 510 research papers in this area. 
Since our focus is more on cross-device interaction with mixed reality for remote collaboration, we first describe the literature of cross-device interaction with mixed reality interfaces. 

\textit{Cross-Device interaction using mixed reality} is an emerging research field in HCI. By leveraging both augmented space and physical devices, mixed reality interfaces allow for both rich visual and physical affordances \cite{muller2015mixed}.
By leveraging augmented visual displays, mixed reality interfaces can enhance the visual affordances and feedback for mobile devices, while keep using tangible devices as a rich tactile and tangible input~\cite{suzuki2020realitysketch, monteiro2023teachable}. 
In prior works, mixed reality cross device interfaces are used for tasks such as extending a device's screen space \cite{baillard2017multi}, interacting with data visualizations~\cite{chulpongsatorn2023holotouch}, or working with 3D models \cite{wu2020megereality}.
BISHARE \cite{zhu2020bishare} provides a nice summary of cross-device interactions with mixed reality interfaces. 
In their paper, they synthesize existing works into ten element design space and showed various ways to leverage both holographic displays and mobile devices together. 
This combination of mixed reality headset and mobile device is increasingly popular in specific application areas.
For example, SymbiosisSketch \cite{arora2018symbiosissketch} uses a MR headset and tablet for 3D spatial sketching, DesignAR \cite{reipschlager2019designar} demonstrates an application for 3D CAD modeling, and MARVIS \cite{langner2021marvis} and HoloTouch~\cite{chulpongsatorn2023holotouch} explore the combination of holographic rendering and tablet inputs for mixed reality data visualizations. 

However, all of these existing works focus on cross-device interaction for \textit{co-located} settings. For example, in all of the above examples, the users and devices need to be located in the same physical space. 
In contrast, this paper focuses on the exploration of how mixed reality interfaces can be applied and how they can augment \textit{remote} collaborative cross-device interaction. 

\subsection{Cross-Device Interaction for Remote Collaboration}
While the majority of existing cross-device interaction research focuses on \textit{co-located} spaces \cite{brudy2019cross}, there are a few research that have explored remote collaboration settings.
For example, Webstrates \cite{klokmose2015webstrates} presents a real-time explorable and shareable dynamic media system that uses a custom web server for online collaboration.
ReticularSpaces \cite{bardram2012reticularspaces} explores the use of activity-based computing to support physically distributed and collaborative smart space that can be used for unified interaction between applications and documents.
Other works have also explored remote collaboration with multiple people and multiple devices \cite{kraut2002understanding, homaeian2021joint, neumayr2018domino, wenzel2020prototyper}.

The main limitation and challenge for cross-device interaction for remote collaboration is the lack of visual affordances. 
For example, a user cannot see how their remote collaborators physically manipulate the device nor the content on it.
Through this work, we open up a new opportunity for cross-device interaction in remote settings by leveraging the real-time holographic rendering to provide rich visual affordances and interactions.

\subsection{Remote Collaboration with Mixed Reality}
Augmented remote collaboration is one of the key application areas for mixed reality technologies. 
In contrast to existing screen-based interfaces, immersive mixed reality interfaces enables rich and expressive spatial collaboration that can provide many benefits. 
For example, Holoportation \cite{orts2016holoportation} demonstrates how the spatial rendering of remote users and objects can enrich remote communication and collaboration.
Three's company \cite{tang2010three} renders a silhouette of the remote user's hands for collaborative work.
These holographic renderings provide unique affordances for spatial information.
A variety of works \cite{lindlbauer2018remixed, harrison2011omnitouch, jones2020vroom} 
that use spatial projection mapping, mobile augmented reality, and mixed reality headsets have shown exciting opportunities for mixed reality and remote collaboration. On the other hand, 
Physical Telepresence \cite{leithinger2014physical}, HoloBots~\cite{ihara2023holobots}, and ChameleonControl~\cite{faridan2023chameleoncontrol} demonstrate remote collaboration through shared physical interaction. 

While many existing works focus on holographic rendering and embodiments of \textit{people and avatars}, the representation of \textit{remote devices} in mixed reality space still remains largely unexplored.
In this work, we show how remote collaboration can be augmented for mixed reality interactions through the holographic rendering of the device's motion and content.
By combining holographic rendering, cross-device interaction, and remote collaboration, we open up new opportunities for HCI and CSCW research.

\section{Holographic Cross-Device Interaction}
This section introduces \textbf{\textit{holographic cross-device interaction}}, a new approach to enabling \textit{remote} and \textit{collaborative} cross-device interaction.
We first outline the overview of the concept, then describe unique benefits of holographic cross device interaction. 
We also present an interaction design space and a the taxonomy of the configuration. 

\begin{figure}[h!]
\centering
\includegraphics[width=\linewidth]{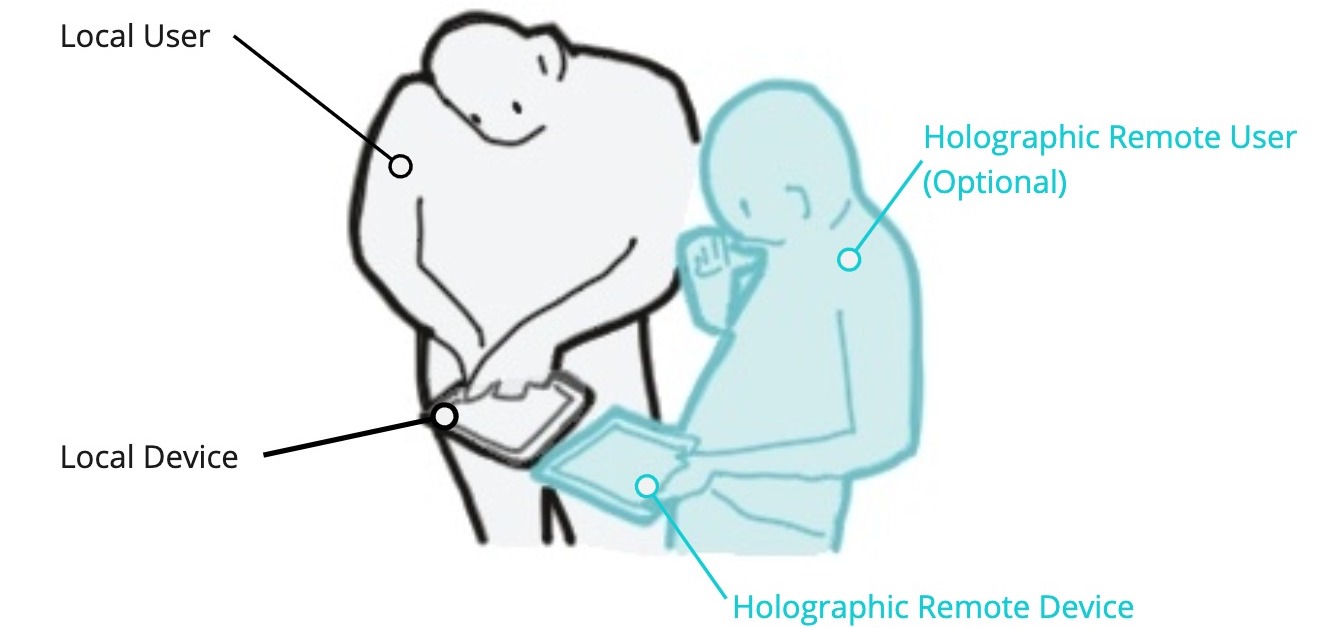}
\caption{General overview of  holographic cross-device interaction. This system relies on three fundamental parts: the local user, the local physical device, and a holographic rendering of the remote device. The remote user can also be rendered to further support the interactions.}
\label{fig:concept}
\end{figure}

\begin{figure*}[h!]
\centering
\includegraphics[width=\textwidth]{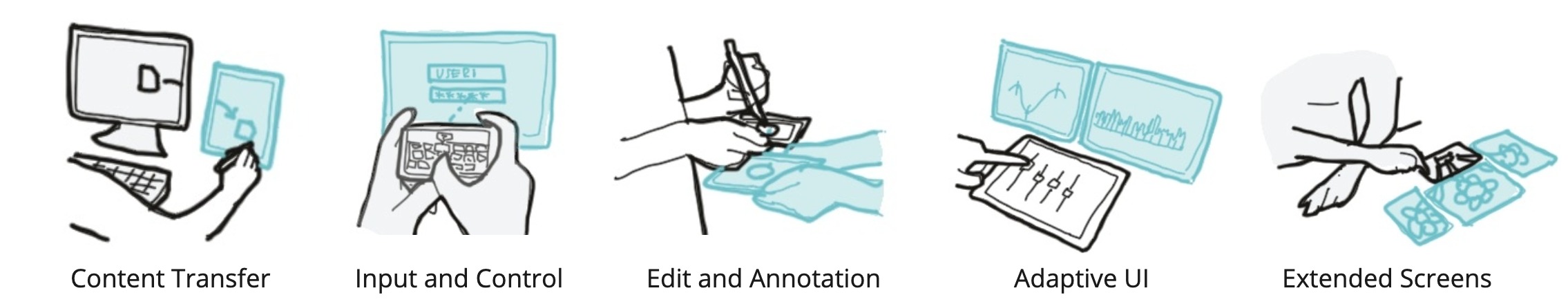}
\caption{Different use cases for the holographic cross-device interaction.}
\label{fig:use-cases}
\end{figure*}

\subsection{Concept and Definition}
Holographic cross-device interaction is an approach to enable \textit{interactions between non co-located devices through holographic rendering of remote devices}.
The remote device's form, motion, and content can be spatially rendered in real-time by leveraging mixed reality displays.
For example, if a remote user moves or rotates the device, the holographic rendering can synchronize the motion as it happens in real-time. 
Moreover, a local user can see the current state and content of the remote device  through a rendering of the screen in mixed reality space. Without the hologram, there would be no visible device to interact with. 
The visual cues of the remote device contextualize it in the local space and allows for cross-device interaction to make logical sense. 
In this way, users can interact with remote devices as if they were all co-located in the same place. 

An embodiment of the remote user can also be rendered in the holographic space as an accompanying visual. By capturing the remote user's hands, face, or even their full body, coordination and collaboration between physically distant users can become more intuitive \cite{tang2010three}.
 
The integration of physical devices, as opposed to a fully holographic system, alleviate the short comings of mixed reality interactions and preserves the benefits of tangible interactions. For example, mid-air interactions are generally imprecise and can be taxing on the user \cite{hincapie2014consumed}. A physical device can provide more comfortable and reliable interactions such as typing on a laptop or sketching with a pen on a tablet. Moreover, users can bring devices they prefer into collaborations, which allows for more varied form factors with their own advantages (such as how precise a mouse is or how natural a pen-display can feel). 

\subsection{Unique Benefits}
There are three major benefits to be gained with holographic cross devices interactions. 

\subsubsection{\textbf{Spatial Visualization of Interaction and Motion}} 
Holographic cross-device interaction can augment a remote device by leveraging the rich visual affordances of mixed reality space for user interaction. 
In non-holographic settings, a local user can only see the content of a remote device on their screen (such as screen sharing on zoom).
With holographic rendering, a remote device can be situated in  physical space and users can see its motion and interact with it through the hologram. This is also true for the rendering of  remote users, which allows for collaborators to spatially interact and coordinate with each other and all of their devices in the space.

\subsubsection{\textbf{Rich Visual Affordances for Intermediate Transition}}
Mixed reality can provide better transition and contextual awareness by visualizing the \textit{intermediate state} of interaction between two or more device. 
For example, this could be a beam pointing from one device to another, an animatation to indicate transition when transfering files, or visual linkages to show how content from different collaborators are interconnected.  
In this way, holographic rendering can \textit{``fill the gap''} between two devices and provide rich visual feedback, which is not possible without mixed reality rendering.

\subsubsection{\textbf{Dynamic and Fluid Configuration}}
In the existing cross-device interaction space, the configurations of devices such as screen position and size are mostly limited due to physical constraints. 
With holographic cross-device interaction, these configuration can be flexible. 
For example, a user can dynamically change the display size of their collaborator's devices regardless of their physical factors. 
In addition, the user can also easily snap or align the remote display as they see fit. 
This dynamic and fluid configuration opens up room for new types of cross-device interactions which has been unexplored thus far.


\subsection{Interaction and Use Cases}
In this section, we discuss the use cases and interaction design space enabled by holographic cross-device interaction. 
Figure \ref{fig:use-cases} summarizes the following five different types of interactions. 

\subsubsection{\textbf{Content Transfer}}
Holographic cross-device interaction allows the user to seamlessly transfer content, data, and information with remote users. 
This is the common use case for existing cross-device interaction \cite{brudy2019cross}, but holographic cross-device interaction allows content transfer in remote collaboration settings to be more spatial and visible. 
For example, figure \ref{fig:use-cases} (content transfer) illustrates how a user can share a file to a different user with a drag-and-drop interaction directly onto the remote device. 
Since the local user can see the remote device as a hologram, it is more natural to transfer these files through screen or gestural interactions.

\subsubsection{\textbf{Input and Control}}
A local device can be used to make input on the remote device or control it in some way. 
When interacting with mixed reality objects, tangible input and control is appreciated for accurate input modality \cite{zhu2020bishare}. 
The user can use their smartphone, tablet, mouse, or keyboard to remotely control and manipulate the other user's device.
For example, figure \ref{fig:use-cases} (input and control) shows how the user can type a text on their smartphone which is being entered in the holographic device.

\subsubsection{\textbf{Content Edit and Annotation}}
In addition to input and control, holographic cross device interaction allows users to edit and annotate content simultaneously. Similar to editing a document online, users can see changes their collaborators make to the document in real-time. 
The difference is that when utilizing this technology the local user can see a holographic version of the remote user and their device while they work as shown in Figure \ref{fig:use-cases} (edit and annotation). This allows for a smooth interaction and coordination as it lets users collaborate as if they are working in the same room. It is also possible for the local user to utilize their physical pen to interact with a remote device through its hologram, which creates a more seamless work space.

\subsubsection{\textbf{Adaptive and Distributed UI}}
In existing cross-device interactions, each device can serves different smaller functionalities of a whole interface (eg. \cite{schmidt2012cross, kubo2017exploring}).
Similarly, holographic cross-device interaction allows elements of the interface to be distributed across different devices. For example, figure \ref{fig:use-cases} (Adaptive UI) shows that the user can collaboratively compose a music through a remotely distributed user interface.
In this way, both local and remote user can create and change the music composition through their synchronously connected user interfaces. 
Again, tangible interactions enabled by the tablet allow for more accurate manipulation than the holographic interactions. 

\subsubsection{\textbf{Extended Screens}}
Holographic cross-device interaction allows the content from one device to be extended and rendered beyond its physical screen space.
The content can be spatially synchronized for all the devices such that they can share the same large content.  
These extended screens are often useful for collaborating on large content that cannot be captured in one device (such as data analysis or whiteboards) or shared experiences (such as a multi-player video game). 

\subsection{Configuration Design Space}
There are several dimensions for the configuration design. These describe how the devices can be set up for holographic cross-device interaction.  

\subsubsection{\textbf{Input Modalities}}
Input modalities describe how the user can interact with the system. Theoretically, there can be as many input modalities as there are devices in the work space. During our research we focused our exploration around common modalities. First, we looked at \textbf{touch} interaction as smart phones and tablets are arguably the most common devices that are used in cross device literature. 
Another is \textbf{gesture} interaction, which is a common form of interaction in both mixed reality and cross device space. We define gesture as manipulation of devices in the space, such as tilting and re-positioning. Lastly, we also considered a \textbf{keyboard and mouse} interaction, as most work space would include a desktop setup.  
    
\subsubsection{\textbf{Arrangement}}
Arrangement describes the spatial configuration between devices, local or remote. These arrangements can be: (1) \textbf{separated} where there is a considerable distance between the two devices and they are only passively interacting with each other at most; (2) \textbf{side-by-side} where the screens are right next to, or in close proximity of, each other such that the user can compare the two screens; or (3) \textbf{overlap} where a device is overlaid on top of the other and the two devices can be treated as one. 

\subsubsection{\textbf{Form}}
Form consists of three variables that describe how the devices exist in the 3-dimensional mixed reality environment. Each of the axis can be leveraged in different ways and interaction between two devices can be parsed from their relative parameters. 

First is the \textbf{position} of the device, which is its location in the space. The relative horizontal and vertical position between two devices can be used for interactions such panning a canvas or scrolling a web page. The depth axis can be used for interactions such as zooming into a picture or aggregating levels of detail on a data visualization. The general distance between two devices can be used for snapping and bumping interactions. 

Second is their \textbf{rotation} which describes the angle of the device about itself. Roll axis can be used for interactions such as animated pouring of objects or to control the sharing of information with another device. The pitch and yaw axis can be used for inspecting and distorting content. 

Last is \textbf{scale} that describes the device's size in the space. In existing cross-device literature, physically smaller device are often used to support interaction with larger device (eg. \cite{neate2017cross, xu2021cross, houben2015watchconnect}). This kind of interaction can be done regardless of the device's physical size in mixed reality because their scale can be virtually adjusted. Moreover, distorting horizontal and vertical scale can be used for interactions such as resizing an image on another device. The depth scale of the device can be use for things such as aggregating the levels of detail on data visualizations, or to control the amount of devices that are allowed to overlap it.   

\subsection{Unique Affordances and Interaction Techniques}
In contrast to existing cross-device interactions, holographic cross-device interaction enables unique visual affordances and interaction techniques that cannot be easily achieved otherwise. The limitations on co-located systems are largely attributed to the constraints of physical reality.
Since holographic cross-device interaction leverages mixed reality interfaces, it is much easier to dynamically change and program the configuration of the devices. 
In this section, we discuss how we can leverage this capability to achieve unique interactions for remote collaboration. 

\begin{figure*}[h!]
\centering
\includegraphics[width=\textwidth]{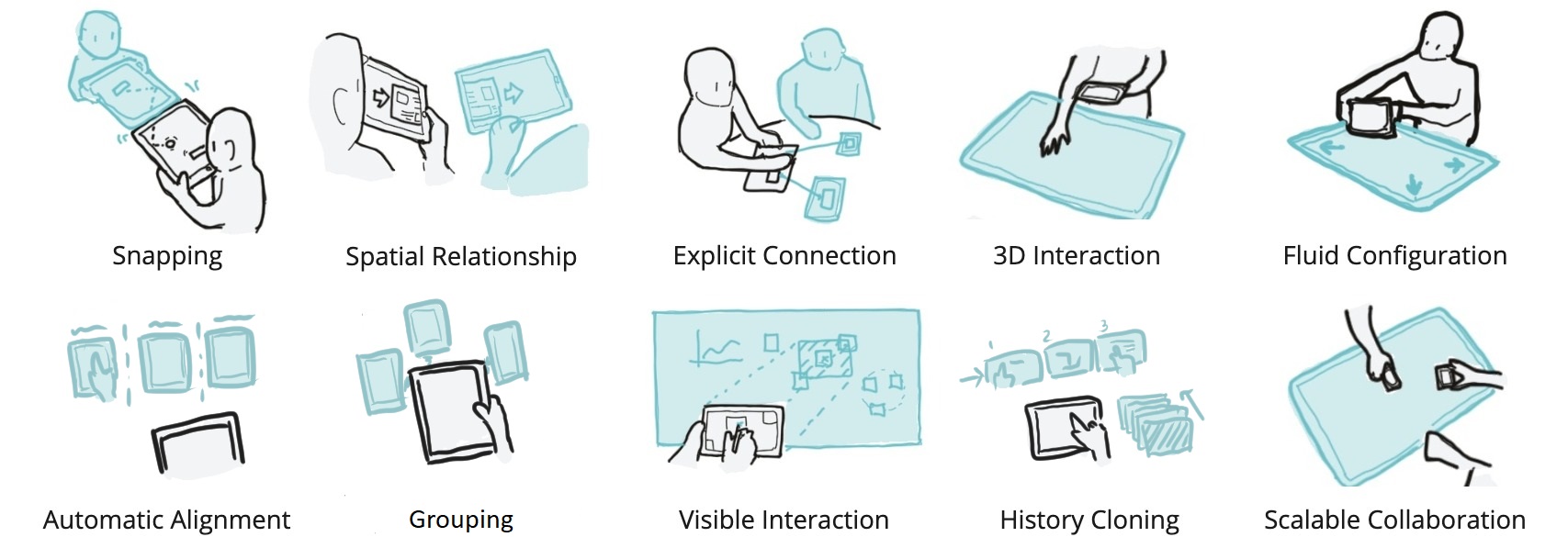}
\caption{Different unique interactions and layouts from the perspective of the local user.}
\label{figure:unique-affordance}
\end{figure*}

\subsubsection{\textbf{Snapping}}
Mixed reality interfaces allows for the \textit{dynamic alignment} of multiple devices. 
By leveraging this capability, the user can easily align multiple devices with \textit{snapping} functionality.
For example, when the user plays a multi-player game like hockey, the holographic cross-device interaction can instantly snap several devices together so that two screens can work as an extended screen. 
This snapping configuration can be maintained even if the user moves or shakes the device.
This functionality allows for easy alignment and configuration in many use cases such as drag-and-drop content transfer or distributed user interfaces.

\subsubsection{\textbf{Automatic Alignment}}
When collaborating with multiple people, the user can not only automatically align screens, but also sort them based on a specific criteria.
For example, when a school teacher needs to oversee a programming exercise, the teacher can automatically sort the student's holographic devices based on their progress to see which students would need some help (similar to Codeopticon \cite{guo2015codeopticon}).
Another example is in collaborative writing. The user can easily align their co-author's devices based on the page they're working on to see how the document would flow. 

\subsubsection{\textbf{Grouping}} 
The user can collect one or more holographic devices into a group to perform interactions on multiple devices at once. For example, if the user needs to physically relocate, they can group the holographic devices to their physical device and move around with them. They user may also share information to multiple devices at once by bumping their physical device onto the group. 

\subsubsection{\textbf{Spatial Relationship}}
The rendering of holographic devices provides visual context and allows for spatial relationships between devices to be leveraged for various interactions.
For example, similar to existing co-located cross-device interaction, the user can transfer their files or content by swiping or "pouring" it to another device.
Similarly, \textbf{proxemic interactions} can be leverage  in remote collaboration settings.
For example, it is possible to dynamically adjust content based on the distance between two devices, which is useful for natural interaction \cite{greenberg2011proxemic} or data analysis and visualization \cite{chulpongsatorn2020exploring}.

\subsubsection{\textbf{3D Interaction}}
Since the remote device is shown as a hologram, it is also possible to \textit{"intersect" or "overlay"} a physical device with a holographic device. 
For example, when a remote user is looking at a 3D sculpture on their screen, a local user can cross their physical device into the holographic device to see the cross-section of the 3D model. Similarly, the user can also look at different layers by moving their devices around inside the holographic device. 
For example, by spatially moving the smartphone, the user can see different levels of human anatomy  or different layer of data visualization.
In this way, the 3D model can be interactively co-explored and annotated.

\subsubsection{\textbf{Fluid Configuration}}
The form variables of the hologram can be dynamically change to adapt to various work space configurations without disrupting the remote collaborator's workflow. The device's position can be changed by dragging the center of the device around, the rotation can be controlled by dragging a side of the hologram in an arc, and the scale can be adjusted by grabbing the corner of the hologram and dragging it outward. For example, this can be used to adjust a smaller virtual device to match with a larger physical device.
It is also possible to clone the device or mirror the device, as well as change the device with a non-linear deformation.

\subsubsection{\textbf{Explicit Connection}}
Holographic visual links can be rendered to explicitly connect content that are distributed across different devices. For example, when the content is copied from a local device, the user can see the explicit linking between the original file and cloned file. In this way, mixed reality environment can be leveraged to show visual elements \textit{outside} of the screens. This was not possible with existing (non mixed reality) cross-device interaction as visual rendering was limited to the boundaries of the displays.

\subsubsection{\textbf{Visible Interaction}}
Similarly, holgraphic cross device interaction can also make interactions more visible through the mixed reality space. Particularly, holographic hands, devices, and any extra holographic peripherals can provide a better visual feedback of what the remote user is currently doing. This entails implicit interactions like eye-gaze, subtle gestures, and movements that can support coordination in remote collaboration.
For example, a user can see their collaborator's form, motion, and pen when they write on a large holographic smartboard.
These visual interactions enrich remote collaboration in a similar manner to ClearBoard \cite{ishii1992clearboard}.

\subsubsection{\textbf{History Cloning}}
The digital nature of the content on the screens allows it to be captured and rendered in a mixed reality space to create a visual history of a device.
With this capability, the user may have different snapshots of a collaborator's device to help them keep track of what has been done so far.
Mixed reality interfaces can show these snapshots in a stack view (like Mac OS's time machine interface) or aligned view (such as a grid or a timeline).
The user can easily reverse the state by overlapping the physical device to the specific history snapshots. 

\subsubsection{\textbf{Scalable Collaboration}}
Finally, since holographic cross-device interaction is not limited by physical space and configuration, the system can be scaled to handle large groups of people. This would be useful for online lectures or teaching where the user needs to handle many remote devices simultaneously. By leveraging the above features and affordances, the holographic cross-device interaction allows for the scalable collaboration of remote settings.

\section{System Implementation}

\subsection{Holographic Rendering}
To demonstrate holographic cross-device interaction, we build \system{}, a proof-of-concept prototype using mixed reality headsets and mobile devices. 
We use Microsoft Hololens 2 head-mounted display for holographic rendering. 
The software runs on Microsoft Hololens Edge browser using WebXR (A-Frame/Three.js) framework. 
All of the screens and their content is rendered as HTML Canvas element.
The canvas element is shown as a canvas texture of a plane geometry element in Three.js. 

\subsection{Content Tracking}
The HTML canvas element is synchronized with the remote device through WebSocket API. DOM attributes are being streamed to the websocket whenever the user explicitly or implicitly interacts with content, such as by drag-and-drop or by physically moving aroud in the space.
These attributes includes but are not limited to: position, color, scale, orientation, and Three.mesh.
In this way, the content on a mobile device can be synchronized with the holograms on the head mounted display. 

\subsection{Device Tracking}
The position and orientation of the mobile device is also captured through WebXR world tracking. We use an off-the-shelf SLAM-based algorithm (8th Wall) for world tracking on the mobile device. In this way, we can capture its current orientation and position in the 3D environment. An HTML document is rendered on top to make the tracking invisible. By sending this captured data to the Hololens via WebSocket and a Node.js server, the device motion rendering can also be synchronized in real-time.

\subsection{User Tracking}
At its core, Holographic cross-device interaction only needs to keep track of the devices and their content. However real-time rendering of the remote user enriches the interactions between users and devices \cite{tang2010three}. 
To this end, we also implement holoportation functionality with 3D video capturing. 
First, we use a Microsoft Azure Kinect depth camera to capture the RGBD information on a Windows 10 machine with 16 GB ram, Nvidia GTX 1060 GPU, and Intel Core i5 CPU. 
We run a Kinect application built on Electron and send its  capture data through WebSocket over the local network. A client can connect to the server and view the point cloud render of the person in 3D Space using A-Frame.
We initially tried to send all of the array elements to the Hololens 2, but the size was much too large and the network latency significantly affected the performance of the client (to about 1 FPS). 
To improve the latency for real-time communication, we reduce the size of array by sampling only every other element in it. We also use the Web Worker API to help with data transmission. This helps improve performance as point cloud data is much too large for WebSocket to send without any pre-processing.
With this simple reduction pipeline, we were able to increase the performance to around 20 FPS, depending on the device the client is running on. 
The latency of the data capturing and rendering of the depth information is approximately 13ms.

We also support synchronized hand tracking over websocket. We use AR.js hand tracking system and send its underlying Three.js object 3D over the web socket as a JSON. This is parsed back into a three JS geometry and mesh, which is then used to reconstruct and render the remote user's hand on the local client. To enable interactions, we parse the position of the reconstructed hand by multiplying the position attribute of its root bone and multiplying that by its scale attribute. 

\subsection{Rapid Prototyping Software Environment}
Based on the above software and hardware framework, we have a basic capturing and rendering pipeline that we leveraged for a rapid prototyping environment using HTML Canvas. In this section we demonstrate our design space  by  implementing an example for each of the use cases we described in section 3.3 using some of the interactions techniques that we listed in section 3.5. We tried to use similar interaction techniques throughout to show how they can be adapted for different use cases. These interactions are governed by various aspects of the configuration design space we discussed in section 3.4.

\subsubsection{\textbf{Content Transfer}}
For content transfer, we used two smartphones that are connected over a websocket. The content of the local device is a simple HTML document with a block of text on it, and the remote device is an empty page. Users can \textbf{visibly} bump the two devices together and send a copy of the text block to the remote device by \textbf{spatially} moving them towards each other. We detect the bump through their position in 3D space. The state of the two devices are not synchronized and the text can be dragged around independently on each screen. The remote device's state is rendered on the Hololens client and the changes are also reflected. 

\begin{figure}[h!]
\centering
\includegraphics[width=\linewidth]{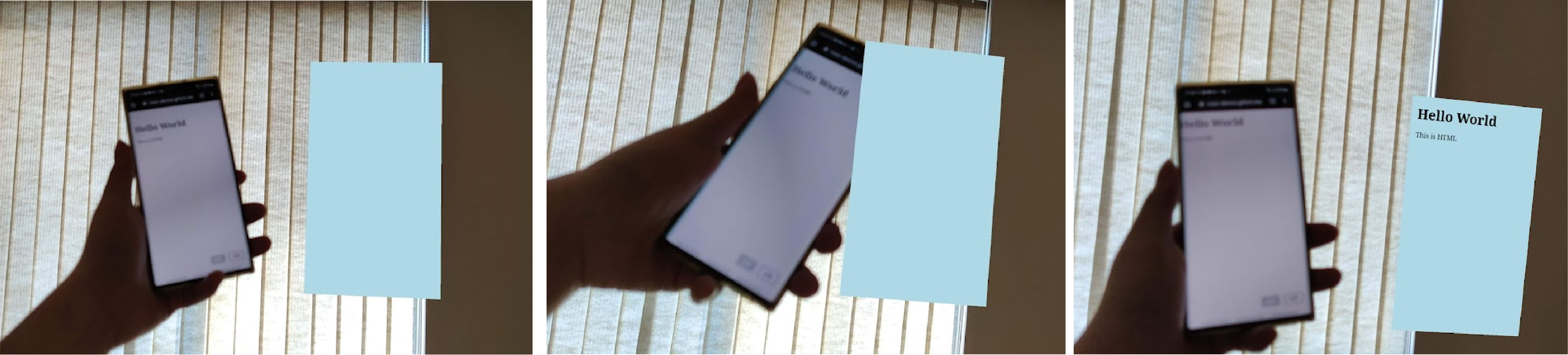}
\caption{User can use the bump gesture to transfer content between devices. The local user moves their device towards the hologram (left). Once the the local phone touches the hologram (middle) the content will be transferred (right).}
\label{figure:implement-bump}
\end{figure}

\subsubsection{\textbf{Input and Control}}
We demonstrated input and control through an implementation of the possession interaction. That is, user can \textbf{spatially}and \textbf{visibly} overlap their physical device on top of the remote device. By holding that position for a moment, the local user can take control of the remote device through their physical device. This is achieved by reading the two devices position. A one second timer is started when the distance between the center of the two devices are very short (< 5cm). The timer is disrupted if the distance becomes larger than the requirement. Once the timer runs out, the hologram of the device disappears, and the content of the remote device is rendered on the local device. The state of the the two devices are synchronized during the possession. The possession can be ended via an on screen button that is available on both devices. 

\begin{figure}[h!]
\centering
\includegraphics[width=\linewidth]{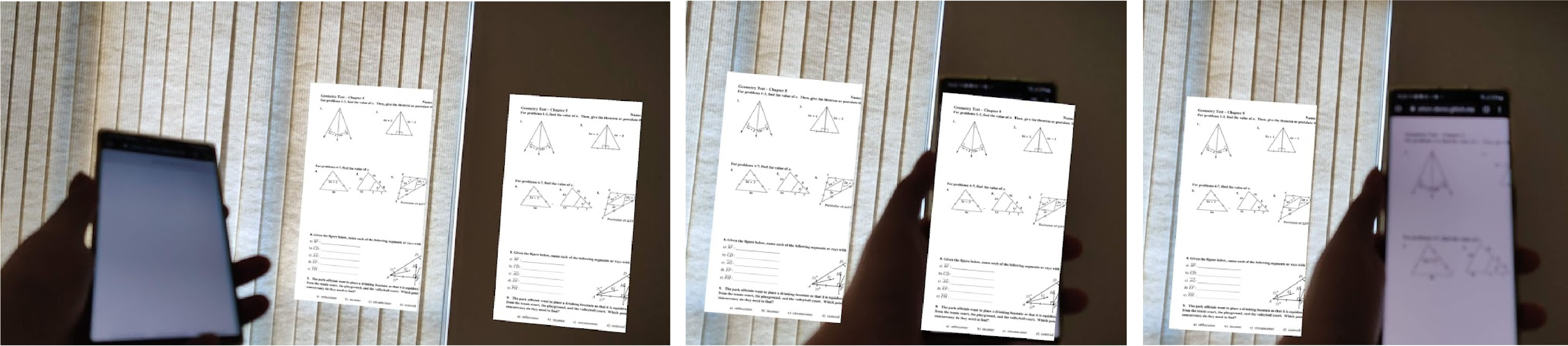}
\caption{ User are able to possess a remote device by overlaying the hologram. One local device and 2 remote devices are displayed (left). The user can move towards and overlap the physical device with one of the hologram (middle). The hologram disappears after a moment and the content is rendered onto the local device (right) }
\label{figure:implement-possess}
\end{figure}

\subsubsection{\textbf{content edit and annotation}}
We demonstrate the content edit and annotation use case by possesing the remote device and writing on it. Remote devices are rendered and spread out in the mixed reality space, each of them has a document on it. The user can \textbf{fluidly} gather the documents and \textbf{automatically sort} them by using a hand gesture. They can then possess a device the same way as described above and type on it using the local device's onscreen keyboard. For hand pose detection, we used Handy Work\footnote{https://github.com/AdaRoseCannon/handy-work}, an open source module for WebXR hand tracking.

\subsubsection{\textbf{Adaptive and Distributed UI}}
We demonstrate dynamic alignment through a multiplayer pong game. An interface that would other wise be on one screen is distributed across multiple users. Users can adapt their interface layout using the \textbf{snapping} functionality. Specifically, a user can attach a holographic device to their physical device by moving close to it. While snapped, the holographic device will translate and rotate with the physical device's motion. User can snap to a different device using the same interaction. The previously snapped device will return to it original position. The user can end the snap interaction via an on-screen button.  

\begin{figure}[h!]
\centering
\includegraphics[width=\linewidth]{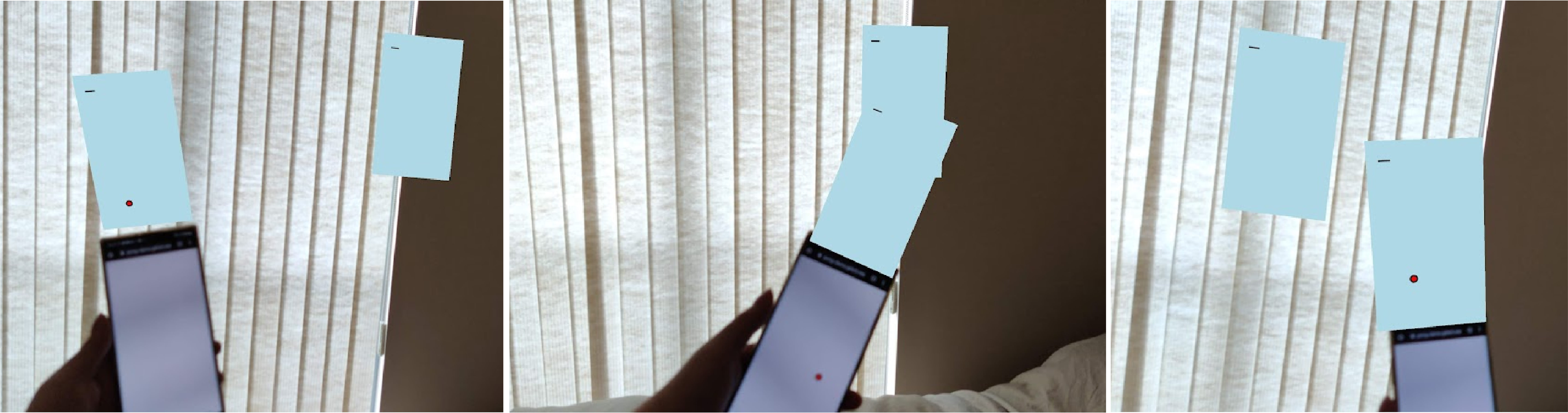}
\caption{We demonstrate a pong game between three devices (1 local and 2 remote) using two features (snapping and extended screen). The left remote device is snapped to the local device, extending the two display space (left). The local user then moves their device toward the right remote device (middle). After a short moment, the right device will snap onto the local device which adapts the over all interface by extending the local screen to a different device instead (right).}
\label{figure:implement-snap}
\end{figure}

\subsubsection{\textbf{Extended Screens}}
For extended screens, we used two smartphones that are connected over websocket. The content of the local device is a simple HTML document with a block of text on it, and the remote device is an empty page. Users can freely move the text between the two devices by dragging them around with touch interactions. When ever a user touches and drags the text block, its position attribute is sent over the websocket to both the holographic client and the remote client. Both clients are \textbf{spatially} aware of where the text block is even when it is not visible, thus creating a extended space. The hand tracking of the remote user is captured on another head mounted display, which is then streamed to the local user's holographic client. This allows the remote user's interaction with the devices to be \textbf{visible} to the local user. The remote device's state is rendered on the Holographic client and the changes are also reflected on the remote device. 

\begin{figure}[h!]
\centering
\includegraphics[width=\linewidth]{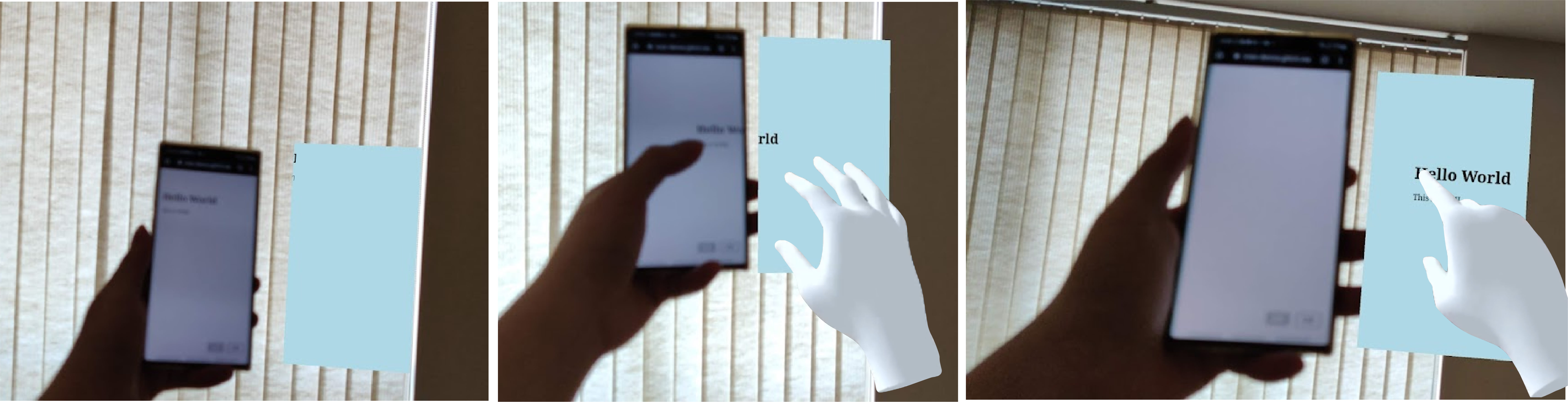}
\caption{Users are able to extended their screen using the hologram. The user would line up their local device with the holographic device (left). Afterwards they drag content from the local device to the remote device as though they were one (middle). The remote user further drags the content into their screen (right).}
\label{figure:implement-extend}
\end{figure}




\section{Application Scenarios}
In this section, we explore possible uses for holographic cross device interaction by describing various application scenarios. None of the users in the following scenarios are co-located.

\subsubsection{\textbf{Brainstorming and Discussion}}
An author is discussing with her editor and artist about their novel. They are standing around a large display that exists in the author's physical space. The author sees all of her collaborators and their tablets through holographic rendering. The tablets provides a better support for writing and drawing as opposed to mid-air interaction with the hologram. During the discussion, the editor writes a post-it note on his tablet and \textbf{transfers} it on to the large display by placing their tablet against the hologram of the large display. Another artist approaches the holographic display, they look at the post-it note, grabs it with their right hand, and \textbf{transfers} it over to their device. The author notices this and she is curious as to what caught the artist's attention. She \textbf{pulls in the hologram} of the artist and their device by using a hand gesture. Through this interaction, the author is able to see what the artist is up to on his tablet without interfering with the artist's work flow.


\subsubsection{\textbf{Collaborative CAD Modeling}}
A 3D modeler is discussing with an engineer about designing a new car model. 
The modeler is showing a 3D car model to the engineer which is displayed on the modeler's desktop machine. Using holographic cross-device interaction, the engineer \textbf{inserts their tablet into the model} to view the cross section on it. The engineer then uses a pen to \textbf{annotate} problematic parts and write down some calculations.
The 3D modeler then turns on \textbf{history cloning} and starts adjusting the model based on the engineer's feedback. They repeat this process a number of times and run into another problem. The 3D modeler then \textbf{places their device on one of the history snapshots} and reverts the changes to that point. 

\subsubsection{Art Education}
An artist is teaching a private digital painting lesson to their student. During, both of their devices are visible to one another. This allows the student to see what the artist is drawing on their end in real time. The opposite is also true, the artist can see what their student is drawing and provide them real-time feedback. To begin, the artist uses hand gestures to pull the reference material that is floating in space towards them. They then drag the reference onto their local device using a mid-air dragging motion with their finger. The artist takes out their physical pen to trace the outline of the reference, then \textbf{bumps} their physical device to the student's holographic device to share the template. The artist also tilts their device above the student's device to \textbf{pour} out the tool's settings for today's lesson. Finally, the artist \textbf{flicks} the reference off of their screen to place it back into the holographic space where the student can easily see it. The lesson can now start.



\subsubsection{Multi-player Gaming}
A student is studying abroad for university but misses playing games back home with his sister. He pulls out his device and connects to his sibling to play remote air hockey together. Both The remote device and the hologram of his sister \textbf{snaps} on to the student's device, \textbf{extending} the display and allowing him to to easily see the state of the game while also maintaining his sister's visual presence. Each device acts as half a field and they are able to move their paddle around through touch interactions on their screen. The hockey puck bounces off the sides and seamlessly travel between each of the device's screens as their match plays out. Once they've had enough of air hockey, the sibling suggests playing Uno. They close the air hockey game, undo the snapping, and they load up Uno. The game starts and their devices now act as their personal hands, hidden from the other. The student grabs a card out of his device and slams it into the floor. The card lays flat in mixed reality space where his sister can see it. The sister takes her turn and does the same. They continue this exchange until they are tired and call it quits.  



\subsubsection{Remote Classroom}
A teacher is giving a lesson on introductory programming to a class of fifteen students. He is able to see a thumbnail size hologram of his students' screens in a straight line formation above his device. After assigning an exercise, the teacher tap one of the screens to enlarge it so that he can see how the student is doing. He then switches out the screen that is in focus using a mid-air \textbf{scrolling} gesture to cycle through his students. While he is doing that, some of his students request assistance. The holographic screens on the teacher side automatically \textbf{reshuffle} and align based on whom requested help first. The teacher looks at the students work and sees where the problem is. He then drags a holographic line originating from his keyboard to the student's holographic device to establish an \textbf{input control}. He then helps debug the student's code by typing on his keyboard. The student is able to see the feedback right away which simulates the experience of being co-located. During this, the teacher notices that some of the other students are also stuck on the same questions. So, he taps each of them to make a group selection, then drags them out so that they can see what is going on. After the class is over, each of the student \textbf{flicks} their assignment in the holographic space before they leave. The teacher then uses a hand gesture to collect them into a grid and brings it towards him. Finally, he starts marking them into the night. 




\subsubsection{Collaborative Document Editing}
A researcher is collaborating on a document with his co-authors. All of his collaborator's screens are scattered around in front of him as holograms. As he scrolls to different part of the text, he can see a holographic line that connects some of the co-author's device to different blocks of text on his screen. The \textbf{visual link} tells the researcher who's currently working on that section and it dynamically pops in and out as blocks of text enters and leave the screen. The researcher can see the changes being made in real time and pulls out his stylus to \textbf{annotate and highlight} portions of the text on his device. Once he is done, he can send his notes to that particular co-author by \textbf{air-tapping the visual link}. In this way, the researcher can suggest edits and give feed back without cluttering everybody else's space.

\section{Discussions and Future Work}

\subsection{Design focused}
This work marks the first exploration into holographic cross device interactions. As such, we focused our efforts on designs and ideas around the holographic cross device interaction space. That means our prototypes are only partially implemented to demonstrate the design space. Rather, we introduce new design opportunities and lay out some ground work that future researchers can leverage to create more concrete, specialized, and complete systems. We truly believe that holographic cross device interaction will open doors to new and exciting interactive systems. 

\subsection{WebXR Challenges} 
While synchronizing primitives and basic parameters is quite trivial, we quickly learned that synchronizing hand and point cloud captures are much more involved. We were able to create a simple demonstration application, but our implementation largely relies on various libraries that each handles their own moving parts. This ate up a lot of development time for something that was not a core part of our contribution. We found that we are still lacking a solid frame work for online synchronization of mixed reality application. 
\section{Conclusion}

This paper introduces \textbf{\textit{holographic cross-device interaction}}, a new class of \textit{remote} cross-device interaction between a local physical device and a holographically rendered remote device. 
Cross-device interactions enable a rich set of interactions between distributed devices, but most of the existing research focus on \textit{co-located} settings (where the users and devices need to be physically situated together in the same place) to achieve these rich interactions and affordances. 
In contrast, holographic cross-device interaction allows remote interactions between devices by providing rich visual affordances through real-time holographic rendering of the device's motion, content, and interactions on mixed reality displays.
Through holographic rendering, not only can remote devices interact as if they are co-located, but they can also be virtually augmented to further enrich the interactions that are not possible with existing cross-device interactions. 
To demonstrate this idea, we developed \system{}, a prototype system for holographic cross-device interaction using Microsoft Hololens 2. 
This paper contributes by introducing the concept of holographic cross-device interaction, presenting the taxonomy and interaction design space of this new domain, demonstrating a variety of uses through our prototype system and application scenarios.

\balance
\bibliographystyle{ACM-Reference-Format}
\bibliography{references}

\end{document}